# A microscopic cranking model for uni-axial rotation with vanishing collective angular momentum

P. Gulshani

NUTECH Services, 3313 Fenwick Crescent, Mississauga, Ontario, Canada L5L 5N1
Tel. #: 647-975-8233; matlap@bell.net

The rigid-irrotational flow transformation in the previous microscopic cranking model (*MCRM*) for nuclear collective rotation about a single axis and its coupling to intrinsic motion is generalized. This generalization allow us to consider the limit of vanishingly small collective angular velocity and hence collective angular momentum while the collective moment of inertia remains finite. In this limit, the collective flow become large and oppose one another collaborating the vanishing of the collective angular momentum. In this limit, the *MCRM* equation for the angular-momentum constraint on the intrinsic wavefunction becomes identical to that of the conventional cranking model (*CCRM*). In this limit, the *MCRM* Schrodinger equation also becomes identical to that of the *CCRM* with an added irrotational-flow kinetic energy component. In this limit, the time-reversal invariance of the *MCRM* Schrodinger equation is destroyed. The two *MCRM* equations (with no free parameters) are solved for the ground-state rotational band in the $^{20}_{10}Ne$ nucleus for a simple deformed harmonic oscillator potential. The predicted excitation energy and quadrupole moment are close to those observed empirically, with the differences seemingly attributable to the absence, in the model, of pairing correlations at low angular momenta and Coriolis-force induced quasi-particle rotation alignment at higher angular momenta. The ground-state terminal (cut-off) angular momentum is predicted to be 10 instead of 8 observed empirically and predicted by the *CCRM*. It is assumed that pairing correlations would reduce the cut-off angular momentum.

## 1. Introduction

The success [1-25 and references therein] of the self-consistent conventional semi-classical cranking model (*CCRM*) [26,27] in predicting rotational properties and phenomena in deformed nuclei behoves us to seek a microscopic foundation for the model[1]. There have been many attempts to achieve this objective with various levels of success using various methods, approximations and assumptions [2,3,10,16,27-32]. The model assumes that the anisotropic nuclear potential $V$ is rotating at a constant angular frequency $\omega_{cr}$ about $x$ or 1 axis. The model time-dependent Schrodinger equation[2]:

$$i\hbar \frac{\partial}{\partial t}|\Psi_{cr}\rangle = H_{cr}|\Psi_{cr}\rangle \tag{1}$$

where:

$$H_{cr} \equiv \frac{1}{2M}\sum_{n,j=1}^{A,3} p_{nj}^2 + V_{cr}(\vec{r}_n), \qquad \vec{r}_n = R(\omega_{cr}t)\,\vec{r}_n' \tag{2}$$

and $R$ is an orthogonal matrix and $\vec{r}_n'$ is the $n^{\text{th}}$ particle coordinate relative to the rotating frame, is then unitarily transformed to the rotating frame:

$$|\Psi_{cr}\rangle = e^{-i(\omega_{cr}L+E)t/\hbar}|\Phi_{cr}\rangle \tag{3}$$

---

[1] Of-course, there are many other models that have had various degrees of success in predicting collective nuclear properties. These other models are not discussed in this article because this article is concerned only with a microscopic derivation of the *CCRM*.

One then obtains the stationary *CCRM* equation[3]:

$$\bar{H}_{cr}\cdot\bar{\Phi}_{cr} \equiv (H-\omega_{cr}\cdot L)\cdot\bar{\Phi}_{cr} = \bar{E}_{cr}\bar{\Phi}_{cr} \tag{4}$$

where $L$ is $x$-component of the total angular momentum operator. The angular velocity $\omega_{cr}$ is then determined by requiring the expectation of $L$ to have a fixed value $\hbar J$:

$$\hbar J \equiv \langle\Phi_{cr}|L|\Phi_{cr}\rangle \tag{5}$$

The energy $E_{cr}$ in a space-fixed frame is then given by:

$$E_{cr} = \langle\Phi_{cr}|H|\Phi_{cr}\rangle = \langle\Phi_{cr}|(H_{cr}+\omega_{cr}\cdot L)|\Phi_{cr}\rangle = \bar{E}_{cr} + \omega_{cr}\cdot\langle\Phi_{cr}|L|\Phi_{cr}\rangle \tag{6}$$

The effective dynamical moment of inertia $\mathcal{J}_{eff}$ is not an observable and must be deduced from other predicted or measured nuclear properties. A definition of $\mathcal{J}_{eff}$, which is adopted from a rigid-body rotation and is commonly used, is given at each value $J$ by the excitation energy $\Delta E_J$:

$$\frac{2\mathcal{J}_{eff}}{\hbar^2} = \frac{4J-2}{\Delta E_J - \Delta E_{J-2}} \quad (MeV)^{-1} \tag{7}$$

---

[2] Clearly, this time-dependent description of the rotational motion is classical in nature because the c-number parameter $\omega_{cr}$ is not an operator acting on a nucleon probability distribution.

[3] Eq. (4) can also be derived from a variation of the Schrodinger equation subject to energy minimization, with the wavefunction $\Phi_{cr}$ constrained to give a fixed value for the expectation of the angular momentum operator.

$$\Delta E_J \equiv E_J - E_0 \tag{8}$$

In this article, we modify the microscopic cranking model (*MCRM*) derived previously in [33,34] to include a more general rigid-irrotational collective flows and show that the collective angular velocity and hence the collective angular momentum can be reduced to zero, and thereby obtain a Schrodinger equation that is identical to that of the *CCRM* except for a relatively small irrotational-flow Coriolis energy term. In Section 2, we derive the new *MCRM*. In Section 3, we solve the *MCRM* Schrodinger equation for a deformed harmonic oscillator potential, and obtain expressions for the (non-collective) angular velocities associated with the angular momentum and shear operators, excitation energy, and quadrupole moment. In Section 4, we apply the model to the nucleus $^{20}_{10}Ne$ and compare the predictions with those of the *CCRM* and experiment. Section 5 present concluding remarks.

## 2. Derivation of microscopic rigid-irrotational flow cranking model with zero collective angular momentum

As in [33,34], the *MCRM* is derived by transforming the nuclear stationary Schrodinger equation using the collective rotation-intrinsic product wavefunction for a rotation about the *x* or 1 axis (as in [35])[4]:

$$\Psi = G(\theta) \cdot \Phi(x_{nj}) \tag{9}$$

where $\theta(x_{nj})$ is the collective-rotation angle and is a function of the space-fixed nucleon co-ordinate $x_{nj}$ ($n=1,...,A$; $j=1,2,3,$ where $A$ is the nuclear mass number) are the space-fixed nucleon co-ordinates. The rotation angle $\theta$ defines the orientation in space of the anisotropic particle distribution (such as quadrupole distribution) described by the intrinsic wavefunction $\Phi$, which is also a function of the space-fixed particle co-ordinates[5]. Applying $\frac{\partial}{\partial x_{nj}}$ and $\frac{\partial^2}{\partial x_{nj}^2}$ to $\Psi$ in Eq. (9), we obtain:

$$\frac{\partial \Psi}{\partial x_{nj}} = \frac{\partial G}{\partial x_{nj}} \Phi + G \frac{\partial \Phi}{\partial x_{nj}} \tag{10}$$

---

[4] In this article, we try as much as possible not to use the phrase "rotating frame", which is a classical-mechanic concept of a frame rotating with a well defined orientation angle and angular velocity as in the *CCRM*.
Note that $\theta$ depends on the spatial distribution of the nucleons and it is not explicitly a function of the nucleon spin. However, since the nucleon spatial distribution is determined by the intrinsic wavefunction $\Phi$, which depends on the nucleon spins, $\theta$ depends indirectly on the spin.
The restriction of the rotation to one spatial dimension is of classical nature but it is adopted here from the conventional cranking model because the objective here is to drive a quantum mechanical analogue of the *CCRM*. This classical feature will be removed when the microscopic model is generalized to 3-D rotation.
The *MCRM* is valid for any nuclear interaction and for a system of fermions or bosons, depending on whether the intrinsic wavefunction is anti-symmetrized or symmetrized respectively.
[5] Note that we do not use any relative co-ordinates for $\Phi$ or anywhere else in the analysis in this article.

$$\frac{\partial^2 \Psi}{\partial x_{nj}^2} = \frac{\partial^2 G}{\partial x_{nj}^2} \Phi + 2 \frac{\partial G}{\partial x_{nj}} \cdot \frac{\partial \Phi}{\partial x_{nj}} + G \frac{\partial^2 \Phi}{\partial x_{nj}^2}$$
$$= \Phi \frac{\partial^2 \theta}{\partial x_{nj}^2} \cdot \frac{dG}{d\theta} + \Phi \frac{\partial \theta}{\partial x_{nj}} \cdot \frac{\partial \theta}{\partial x_{nj}} \cdot \frac{d^2 G}{d\theta^2} + 2 \frac{\partial \theta}{\partial x_{nj}} \cdot \frac{dG}{d\theta} \cdot \frac{\partial \Phi}{\partial x_{nj}} + G \frac{\partial^2 \Phi}{\partial x_{nj}^2} \quad (11)$$

Substituting Eq. (11) into the stationary Schrodinger equation:

$$H \cdot \Psi \equiv \left( \frac{1}{2M} \sum_{n,j=1}^{A,3} p_{nj}^2 + V \right) \cdot \Psi = E \cdot \Psi \quad (12)$$

where $M$ is the nucleon mass and $V$ is an arbitrary nuclear interaction, we obtain:

$$G \cdot H \cdot \Phi - \frac{\hbar^2}{M} \cdot \sum_{n,j} \frac{\partial \theta}{\partial x_{nj}} \cdot \frac{dG}{d\theta} \cdot \frac{\partial \Phi}{\partial x_{nj}} - \frac{\hbar^2}{2M} \cdot \Phi \cdot \sum_{n,j} \frac{\partial^2 \theta}{\partial x_{nj}^2} \cdot \frac{dG}{d\theta}$$
$$- \frac{\hbar^2}{2M} \cdot \Phi \cdot \sum_{nj} \frac{\partial \theta}{\partial x_{nj}} \cdot \frac{\partial \theta}{\partial x_{nj}} \cdot \frac{d^2 G}{d\theta^2} = E \cdot \Phi \quad (13)$$

We require the orientation $\theta$ of the deformed nuclear nucleon distribution to be defined by the motion of the particles and hence by the angular momentum operator $L$ along $x$ or $1$[6]. Therefore, $\theta$ and $L$ are a canonically conjugate pair, satisfying the commutation relation:

$$[\theta, L] = i\hbar \quad \Rightarrow \quad L \equiv \sum_n (y_n p_{nz} - z_n p_{ny}) = -i\hbar \frac{\partial}{\partial \theta} \quad (14)$$

Substituting Eq. (14) into Eq. (13), we obtain:

$$G \cdot H \cdot \Phi + \frac{1}{M} \cdot \sum_{n,j} \frac{\partial \theta}{\partial x_{nj}} \cdot (L \cdot G) \cdot p_{nj} \cdot \Phi - \frac{i\hbar}{2M} \cdot \Phi \cdot \sum_{n,j} \frac{\partial^2 \theta}{\partial x_{nj}^2} \cdot L \cdot G$$
$$+ \frac{1}{2M} \cdot \Phi \cdot \sum_{nj} \frac{\partial \theta}{\partial x_{nj}} \cdot \frac{\partial \theta}{\partial x_{nj}} \cdot L^2 \cdot G = E \cdot \Phi \quad (15)$$

Next we assume that $G$ is an eigenstate of $L$:

$$L e^{il\theta} = \hbar \gamma e^{i\gamma\theta} \quad (16)$$

where $\hbar\gamma$ is the angular momentum associated with the collective rotation. $\gamma$ is determined later in this section. Substituting Eq. (16) into Eq. (15), we obtain:

---

[6] Note that $L$ can be considered to be the total angular momentum including the particle spin because $\theta$ does not depend explicitly on the spin as discussed in footnote 4.

$$H \cdot \Phi + \frac{\hbar \gamma}{M} \cdot \sum_{n,j} \frac{\partial \theta}{\partial x_{nj}} \cdot p_{nj} \cdot \Phi + \frac{\hbar^2 \gamma^2}{2M} \cdot \sum_{n,j} \frac{\partial \theta}{\partial x_{nj}} \cdot \frac{\partial \theta}{\partial x_{nj}} \cdot \Phi - \frac{i\gamma \hbar^2}{2M} \cdot \Phi \cdot \sum_{n,j} \frac{\partial^2 \theta}{\partial x_{nj}^2} = E \cdot \Phi \quad (17)$$

Physical descriptions of the various terms in Eq. (17) can be found in [33,34]. As in [33,34], we define the rotation angle $\theta$ in terms of the nuclear quadrupole distribution since observation (experimental and theoretical) indicate that nuclear rotational motion is dominated by the quadrupole nucleon distribution (Bohr-Mottelson's quadrupole deformation model and numerous other collective models such as Villars' collective models using quadrupole moment to define the rotation angle are a testament to this fact). In line with this observation, we define the rotation angle $\theta$ to satisfy the relation (for a rotation about $x$ or 1 axis only)[7]:

$$\frac{\partial \theta}{\partial x_{nj}} = \sum_{k=1}^{2} \bar{\chi}_{jk} x_{nk}, \quad \chi_{jk} = 0 \text{ for } j,k \neq 2,3 \quad (18)$$

The real 3x3 matrix $\bar{\chi}$ can be chosen to be a sum of different types of matrices, each describing a different type of physical motion such as quadrupole rigid and irrotational, and non-quadrupole rigid flow regimes described in [35-44]. In this article, we choose $\chi$ to be the sum of an antisymmetric, a symmetric, and an arbitrary matrix so that the non-zero elements of $\chi$ are:

$$\bar{\chi}_{23} \equiv (1+\lambda) \cdot \chi + \chi_3, \text{ and } \bar{\chi}_{32} = -(1-\lambda) \cdot \chi - \chi_2 \quad (19)$$

for arbitrary real parameters $\chi$, $\lambda$, $\chi_2$ and $\chi_3$. Substituting Eq. (19) into $[\theta, L] = i\hbar$ in Eq. (14), we obtain:

$$\chi = -\frac{1+C_+}{\mathcal{J}_+ - \lambda \cdot \mathcal{J}_-} \equiv -\mathcal{J}^{-1} \quad (20)$$

where the intrinsic rigid-flow $\mathcal{J}_+$ and deformation $\mathcal{J}_-$ moments of inertia are defined as:

$$\mathcal{J}_+ \equiv \sum_n (y_n^2 + z_n^2), \quad \mathcal{J}_- \equiv \sum_n (y_n^2 - z_n^2), \quad C_+ \equiv \sum_n (\chi_2 y_n^2 + \chi_3 z_n^2) \quad (21)$$

---

[7] Classically, $\frac{\partial \theta}{\partial x_{nj}}$ may be considered to be the collective component of the particle velocity field, refer to [35] for more detail. For any linear (in Eq. (18)) or other flow prescription for $\theta$, one can prove (using Eqs. (18)-(20) or Stoke's theorem, refer to [35, Eq. (57)]) that, for a system of more than one particle, the mixed second partial derivatives of $\theta$ are discontinuous, i.e., $\vec{\nabla}_n \times \vec{\nabla}_n \theta \neq 0$. This discontinuity seems to be related to the observation that a change $\delta \theta$ in the collective angle $\theta$ corresponds to different sets of changes $\delta \vec{r}_n$ in the particle positions in a multi-particle system. Even for a single particle, $\vec{\nabla} \times \vec{\nabla} \theta \neq 0$ at the coordinate system origin. However, this discontinuity is of no consequence for the analysis presented in this article because no mixed second derivative of $\theta$ appears anywhere in the analysis and all the derived variables are continuous and well behaved.

We now substitute Eqs. (18)-(20) into Eq. (17), and ignore the third term on the left-hand side of Eq. (17) arising from the action of $L$ on $\mathcal{I}_+$ and $\mathcal{I}_-$ (i.e., the term arising from the interaction of rotation with fluctuations in intrinsic nucleon quadrupole distribution) because this term is relatively small, and its expectation over the state $\Phi$ generally vanishes, and such terms are excluded from in the *CCRM* since the angular velocity is a constant in the *CCRM*. For this reason we also replace $\mathcal{I}_+$ and $\mathcal{I}_-$ by their expectation values in the rest of this article as in [33,34]:

$$\mathcal{I}_+^o \equiv \langle \Phi | \mathcal{I}_+ | \Phi \rangle, \quad \mathcal{I}_-^o \equiv \langle \Phi | \mathcal{I}_- | \Phi \rangle, \quad \mathcal{I}^o \equiv \langle \Phi | \mathcal{I} | \Phi \rangle, \quad \omega^o \equiv \frac{\hbar \gamma}{M \mathcal{I}^o} \qquad (22)$$

We then obtain[8]:

$$\left( H + \bar{\omega}_+ \cdot L - \bar{\omega}_- \cdot T \right) \cdot \Phi = \bar{E} \cdot \Phi \qquad (23)$$

where:

$$\frac{\bar{\omega}_+}{\omega^o} \equiv 1 - \mathcal{I}^o \cdot \frac{\chi_2 + \chi_3}{2}, \quad \frac{\bar{\omega}_-}{\omega^o} \equiv -\lambda - \mathcal{I}^o \cdot \frac{\chi_2 - \chi_3}{2}, \quad \omega^o \equiv \frac{\hbar \gamma}{M \mathcal{I}^o} \qquad (24)$$

$$\bar{E} \equiv E - \frac{M \omega^{o2}}{2} \cdot \left[ \left(1+\lambda^2\right) \cdot \mathcal{I}_+^o - 2\lambda \cdot \mathcal{I}_-^o - 2\mathcal{I}^o \cdot \left( C_+ - \lambda \cdot C_- \right) + C \cdot \mathcal{I}^{o2} \right] \equiv E - E_{cf} \qquad (25)$$

$$C_- \equiv \sum_n \left( \chi_2 \cdot \langle y_n^2 \rangle - \chi_3 \cdot \langle z_n^2 \rangle \right), \quad C \equiv \sum_n \left( \chi_2^2 \cdot \langle y_n^2 \rangle + \chi_3^2 \cdot \langle z_n^2 \rangle \right) \qquad (26)$$

$\langle y_n^2 \rangle \equiv \langle \Phi | y_n^2 | \Phi \rangle$ etc., and $T$ is a linear shear operator, generating a linear irrotational flow, and defined by:

$$T \equiv \sum_n \left( y_n p_{nz} + z_n p_{ny} \right) \qquad (27)$$

In Eq. (23), we have ignored the rotation-fluctuation interaction term in (the last term on the left-hand side of ) Eq. (17) because it is small (in fact, its expectation vanishes), and we want correspondence with the *CCRM* where this interaction is not considered. The *CCRM* angular momentum constraint in Eq. (5) then becomes using Eqs. (9) and (16)[9]:

$$\hbar J = \langle \Psi | L | \Psi \rangle = \langle G | L | G \rangle + \langle \Phi | L | \Phi \rangle = \hbar \gamma + \langle \Phi | L | \Phi \rangle \qquad (28)$$

---

[8] Eq. (23) is somewhat similar to that used in [43,44] where a vortex-flow angular momentum given in [39] was phenomenologically added to the *CCRM* as a constraint.

[9] The value of $\gamma$ determined by Eq. (28) may be viewed as an approximation (implied by the *CCRM*) to an integer value of $\gamma$ needed to ensure that $\Psi$ is single-valued function of $\theta$.

## 2.1 Vanishing collective angular momentum

We now show that it is possible to choose $\omega^o$ and hence $\gamma$ in Eq. (24) vanishingly small while $\mathcal{J}^o$ remains finite, and thereby reduce Eq. (28) to the *CCRM* Eq. (5)[10]. First we observe that Eq. (23) can be expressed similarly to the *CCRM* Eq. (4) by using the identity: $\bar{\omega}_+ = -(-\bar{\omega}_+)$. We know that $\omega_{cr}$ in Eq. (4) is a positive quantity. Therefore, $\bar{\omega}_+$ must be a negative quantity. From the frequency solution in [33,34] of Eq. (23) for a deformed harmonic oscillator potential, we know that the product $-\bar{\omega}_+ \cdot \bar{\omega}_-$ appears in this solution. This product causes the frequencies to behave similarly to the frequencies determined from the *CCRM* Eq. (4) only if this product is positive. Therefore, we must require that $\bar{\omega}_-$ be a positive quantity. To ensure that $\bar{\omega}_+$ and $\bar{\omega}_-$ are respectively negative and positive quantities, we conveniently express Eqs. (24) in the following forms (without loss of generality):

$$\frac{\bar{\omega}_+}{\omega^o} \equiv 1 + \frac{\nu}{\omega^o} \cdot \left(\frac{\mathcal{J}^o_+}{\mathcal{J}^o_-}\right)^{n_1}, \quad \frac{\bar{\omega}_-}{\omega^o} \equiv -\lambda + \frac{\mu}{\omega^o} \cdot \left(\frac{\mathcal{J}^o_+}{\mathcal{J}^o_-}\right)^{n_2} \tag{29}$$

where $\nu, \mu > 0$, and $n_1$ and $n_2$ are respectively odd and even integers, noting that $\mathcal{J}^o_+$ and $\mathcal{J}^o_-$ are respectively positive and negative quantities for a prolate shape. In the limit of small $\omega^o$ in Eqs. (29), we obtain:

$$\frac{\bar{\omega}_+}{\bar{\omega}_-} \xrightarrow[\omega^o \to 0]{} \frac{\nu}{\mu} \cdot \left(\frac{\mathcal{J}^o_+}{\mathcal{J}^o_-}\right)^{n_1 - n_2} < 0 \tag{30}$$

Substituting Eqs. (21), (22), (24), and (29) into Eq. (20), we obtain:

$$\mathcal{J}^o \cdot \omega^o \xrightarrow[\omega^o \to 0]{} \mu \cdot \left(\frac{\mathcal{J}^o_+}{\mathcal{J}^o_-}\right)^{n_2} \cdot \mathcal{J}^o_- \cdot \left[\frac{\nu}{\mu} \cdot \left(\frac{\mathcal{J}^o_+}{\mathcal{J}^o_-}\right)^{n_1 - n_2} + 1\right] \tag{31}$$

Eq. (31) shows that $\mathcal{J}^o \cdot \omega^o$ and hence $\gamma$ in Eq. (24) vanish for:

$$\frac{\nu}{\mu} \cdot \left(\frac{\mathcal{J}^o_+}{\mathcal{J}^o_-}\right)^{n_1 - n_2} = -1 \tag{32}$$

Eq. (32) is compatible with the condition in Eq. (30). Substituting Eq. (32) into Eq. (30), we obtain, in the limit of small $\omega^o$,:

---

[10] Note that in the *MCRM* Schrodinger Eq. (23), we identify $\bar{\omega}_+$ with $-\omega_{cr}$ in the *CCRM* Schrodinger Eq. (4), and distinguish it from the collective angular velocity $\omega$, whereas in the CCRM $\omega_{cr}$ is the angular velocity of the rotating frame.

$$\bar{\omega}_- = -\bar{\omega}_+ \qquad (33)$$

Comparison of Eqs. (24) and (29) shows that, in the limit of small $\omega^o$, $\bar{\omega}_+$ becomes independent of $\omega^o$, i.e., as the collective angular velocity $\omega^o$ (of the rotating frame) tends to zero, the angular velocity $\bar{\omega}_+$ remains finite and is determined by the angular momentum constraint in Eq. (28) with zero collective angular momentum $\hbar\gamma$. Therefore, the *MCRM* distinguishes between the collective angular velocity and the angular velocity associated the Coriolis energy term $\bar{\omega}_+ \cdot (L-T)$ in the Schrodinger Eq. (23) for the intrinsic wavefunction.

From the definitions in Eqs. (24) and (29) we obtain, in the limit of small $\omega^o$, (using Eq. (32)):

$$-\mathcal{J}^o \cdot \frac{\chi_2 + \chi_3}{2} \bigg/ -\mathcal{J}^o \cdot \frac{\chi_2 - \chi_3}{2} = \frac{\nu}{\omega}\left(\frac{\mathcal{J}^o_+}{\mathcal{J}^o_-}\right)^{n_1} \bigg/ \frac{\mu}{\omega}\left(\frac{\mathcal{J}^o_+}{\mathcal{J}^o_-}\right)^{n_2} = \frac{\nu}{\mu}\left(\frac{\mathcal{J}^o_+}{\mathcal{J}^o_-}\right)^{n_1 - n_2} = -1 \Rightarrow \chi_2 \xrightarrow[\omega^o \to 0]{} 0 \quad (34)$$

From Eqs. (24) and (29) and result in Eq. (34), we obtain:

$$\mathcal{J}^o \cdot \chi_3 = -\frac{2\nu}{\omega}\left(\frac{\mathcal{J}^o_+}{\mathcal{J}^o_-}\right)^{n_1} = \frac{2\mu}{\omega}\left(\frac{\mathcal{J}^o_+}{\mathcal{J}^o_-}\right)^{n_2} \Rightarrow \chi_3 \xrightarrow[\omega^o \to 0]{} \infty \qquad (35)$$

Substituting the result in Eq. (34) into Eq. (20) and using Eqs. (21) and (22), we obtain:

$$\chi = -\frac{1+C_+}{\mathcal{J}^o_+ - \lambda \cdot \mathcal{J}^o_-} \xrightarrow[\omega^o \to 0]{} -\frac{\sum_n \chi_3 \langle z_n^2 \rangle}{\mathcal{J}^o_+ - \lambda \cdot \mathcal{J}^o_-} = -\frac{1}{2}\frac{\mathcal{J}^o_+ - \mathcal{J}^o_-}{\mathcal{J}^o_+ - \lambda \cdot \mathcal{J}^o_-} \cdot \chi_3 \xrightarrow[\omega^o \to 0]{} -\infty \qquad (36)$$

Substituting Eqs. (34), (35), and (36) into Eq. (19), we obtain:

$$\bar{\chi}_{23} = \frac{1}{2}(1-\lambda)\frac{\mathcal{J}^o_+ + \mathcal{J}^o_-}{\mathcal{J}^o_+ - \lambda \cdot \mathcal{J}^o_-}\cdot\chi_3, \qquad \bar{\chi}_{32} = \frac{1}{2}(1-\lambda)\frac{\mathcal{J}^o_+ - \mathcal{J}^o_-}{\mathcal{J}^o_+ - \lambda \cdot \mathcal{J}^o_-}\cdot\chi_3 \qquad (37)$$

We can express $\bar{\chi}_{23}$ and $\bar{\chi}_{32}$ as: $\bar{\chi}_{23} = \sigma_1 + \sigma_2$, $\bar{\chi}_{32} = \sigma_1 - \sigma_2$ where $\sigma_1$ and $\sigma_2$ are the non-zero elements of an antisymmetric $\Omega^a$ and a symmetric $\Omega^s$ matrix respectively. We then find:

$$\sigma_1 = \frac{1}{2}\cdot\frac{1-\lambda}{\mathcal{J}^o_+ - \lambda \cdot \mathcal{J}^o_-}\cdot\chi_3 \cdot \mathcal{J}^o_- < 0, \qquad \sigma_2 = \frac{1}{2}\cdot\frac{1-\lambda}{\mathcal{J}^o_+ - \lambda \cdot \mathcal{J}^o_-}\cdot\chi_3 \cdot \mathcal{J}^o_+ > 0 \qquad (38)$$

The streamlines described by the matrices $\Omega^a$ and $\Omega^s$ flow in opposing directions, and each of them vanish for the particular choice: $\lambda = 1$, noting that $\lambda$ is an arbitrary parameter with no impact on any other feature of the model. The two opposing flows or currents may be viewed as follows. We may define the collective velocity component $\vec{v}_n^{\,rot}$ of the $n^{\text{th}}$ nucleon velocity (as in

[35], see also footnote 7) and x-component of the collective angular momentum $L_{coll}$ by:

$$\vec{v}_n^{rot} \equiv a_o \vec{\nabla}_n \theta, \qquad L_{coll} \equiv \sum_n M \cdot a_o \cdot \left( y_n \cdot \frac{\partial \theta}{\partial z_n} - z_n \cdot \frac{\partial \theta}{\partial y_n} \right)$$

Using Eqs. (18) and (37), we then obtain the result:

$$L_{coll} = M \cdot a_o \cdot c_o \left[ \left( \mathcal{I}_+^o - \mathcal{I}_-^o \right) \cdot \sum_n y_n^2 - \left( \mathcal{I}_+^o + \mathcal{I}_-^o \right) \cdot \sum_n z_n^2 - \right]$$

$$= \frac{1}{2} M \cdot a_o \cdot c_o \left[ \left( \mathcal{I}_+^o - \mathcal{I}_-^o \right) \cdot \left( \mathcal{I}_+^o + \mathcal{I}_-^o \right) - \left( \mathcal{I}_+^o + \mathcal{I}_-^o \right) \cdot \left( \mathcal{I}_+^o - \mathcal{I}_-^o \right) \right] = 0$$

where $c_o \equiv \frac{1}{2} \cdot \frac{1-\lambda}{\mathcal{I}_+^o - \lambda \cdot \mathcal{I}_-^o} \cdot \chi_3$. This result implies that the opposing collective flows cancel one another. Substituting Eqs. (31), (32), and (33) into Eqs. (23) and (28), we obtain:

$$\left[ H + \bar{\omega}_+ \cdot (L - T) \right] \cdot \Phi = \bar{E} \cdot \Phi \qquad (39)$$

$$\hbar J = \langle \Phi | L | \Phi \rangle \qquad (40)$$

Eqs. (39) and (40) are identical to the *CCRM* Eqs. (4) and (5), when we identify $\bar{\omega}_+$ with $-\omega_{cr}$, except for the irrotational-flow kinetic energy term $\bar{\omega}_+ \cdot T$ (i.e., irrotational-flow Coriolis energy term) in Eq. (39). We show in Section 4 that the consequences (among others) of this term are higher rotational-band terminal (i.e., cut off) angular momentum, and lower angular velocity and excitation energy than those in the *CCRM*.

### 3. Solution of Eqs. (39) and (40) for deformed oscillator potential

Eq. (39) is similar to Eq. (29) in [33] or Eq. (26) in [34]. Therefore, for a simple deformed harmonic oscillator potential[11]:

$$H = \frac{1}{2M} \cdot \sum_{n,j=1}^{A,3} p_{nj}^2 + \frac{M\omega_1^2}{2} \cdot \sum_n x_n^2 + \frac{M\omega_2^2}{2} \cdot \sum_n y_n^2 + \frac{M\omega_3^2}{2} \cdot \sum_n z_n^2 \qquad (41)$$

the solution of Eq. (39) is readily obtained from that in Eqs (45)-(48) in [33] or Eqs. (35)-(37) in [34]. We then obtain the energy eigenvalue for Eq. (39) in a space-fixed frame:

$$E = \hbar \omega_1 \Sigma_1 + \hbar \alpha_2 \Sigma_2 + \hbar \alpha_3 \Sigma_3 + \frac{M}{2} \bar{\omega}_+^2 \cdot \left[ \mathcal{I}_+^o - \mathcal{I}_-^o \right] \qquad (42)$$

---

[11] The simple choice of the nuclear interaction $V$ in Eq. (41) is made because it yields analytical solutions and hence facilitates identification and explanation of any differences between the predictions of the *MCRM* and *CCRM*. Of-course, subsequent calculations using realistic $V$ need to be performed to realistically quantify the impact of the discrepancies.

where:

$$\alpha_2^2 \equiv \omega_+^2 + \sqrt{\omega_-^4 + 4\bar{\omega}_+^2 \cdot (\omega_+^2 + \omega_-^2)}, \quad \alpha_3^2 \equiv \omega_+^2 - \sqrt{\omega_-^4 + 4\bar{\omega}_+^2 \cdot (\omega_+^2 + \omega_-^2)} \tag{43}$$

$$\omega_+^2 \equiv \frac{\omega_2^2 + \omega_3^2}{2}, \quad \omega_-^2 \equiv \frac{\omega_2^2 - \omega_3^2}{2}, \quad \Sigma_k \equiv \sum_{n_k=0}^{n_{kf}} (n_k + 1/2) \tag{44}$$

where $n_{kf}$ is the number of oscillator quanta in the $k^{th}$ direction at the Fermi surface. Note that, due to the term $\bar{\omega}_+ \cdot T$ in Eq. (39), the frequencies in Eq. (43) differ from those in [33,34] predicted by the *CCRM*:

$$\alpha_{cr2}^2 \equiv \omega_+^2 + \omega_{cr}^2 + \sqrt{\omega_-^4 + 4\omega_{cr}^2 \cdot \omega_+^2}, \quad \alpha_{cr2}^2 \equiv \omega_+^2 + \omega_{cr}^2 - \sqrt{\omega_-^4 + 4\omega_{cr}^2 \cdot \omega_+^2}$$

This difference results in higher rotational-band terminal (i.e., cut off) angular momentum, and lower angular velocity and excitation energy than those in the *CCRM*.

For $H$ in Eq. (41) to approximate a Hartree-Fock mean-field Hamiltonian, we minimize the energy $E$ in Eq. (41) with respect to the frequencies $\omega_k$ ($k = 1, 2, 3$) at a fixed value of $J$ and hence of $\mathcal{J}^o$, $\bar{\omega}_+$, $\mathcal{J}_+^o$, and $\mathcal{J}_-^o$ given by the constraint in Eq. (40), subject to the constant nuclear-quadrupole-volume condition:

$$\langle x^2 \rangle \cdot \langle y^2 \rangle \cdot \langle z^2 \rangle = c_o \tag{45}$$

where $\langle x_k^2 \rangle \equiv \langle \Phi | \sum_n x_{nk}^2 | \Phi \rangle$ ($k = 1, 2, 3$) and $c_o$ is a constant. This minimization yields a self-consistency between the shapes of nuclear equi-potential and equi-density surfaces [5,45-48]. The minimization is performed numerically as in [47,48].

## 4. Model predictions for $^{20}_{10}Ne$

In this section we present the excitation energy ($\Delta E_J$) and quadrupole moment ($Q_o$) predicted by the *MCRM* and *CCRM* for $^{20}_{10}Ne$ ground-state rotational band using the solutions given in Section 3 for the simplest possible interaction, namely the deformed harmonic oscillator potential. This potential is used because it gives an analytic solution and hence facilitates the identification and explanation of the differences between the predictions of the two models. A more realistic potential will be used later to quantify realistically the impact of these differences. The model results are also compared with the measured data. For $^{20}_{10}Ne$, we use the anisotropic-harmonic-oscillator nucleon-occupation configuration $(\Sigma_1, \Sigma_2, \Sigma_3) = (14, 14, 22)$, with the spherical harmonic oscillator frequency $\hbar\omega_o = 35.4 \cdot A^{-1/3}$ *MeV* as in [47,48].

Fig 1 shows that the *MCRM* predicts lower $\Delta E_J$ at and below $J = 4$ and higher $\Delta E_J$ above $J = 4$, particularly at $J = 8$ than observed empirically. This behaviour is assumed to be caused by the neglect of pairing interaction in the model. At low $J$ values, pairing interaction couples pairs of nucleons to zero angular momentum reducing the moment of inertia and hence increasing $\Delta E_J$. At higher $J$ values, Coriolis force tends to align the quasi-particle angular momenta along the rotation eventually breaking up the paired nucleons resulting in higher moment of inertia and hence lower $\Delta E_J$. The *MCRM* predicts lower angular velocity and hence lower $\Delta E_J$ than the *CCRM* because of the presence of the irrotational-flow Coriolis term $\bar{\omega}_+ \cdot T$ in the *MCRM* Eq. (39), which results in cranked frequencies different than those predicted by the *CCRM* as noted in Section 3.

Fig 2 shows that the *MCRM* predicts decreasing $Q_o$ with $J$ as does the *CCRM* and experiment. However, the $Q_o$ predicted by the *MCRM* is consistently higher particularly at $J = 8$, where the band is predicted by the *CCRM* and observed empirically to terminate when the nucleus is predicted by the *CCRM* to become axially symmetric about the rotation axis. On the other hand, the *MCRM* predicts that the nucleus becomes axially symmetric and the band terminates at $J = 10$. This difference between terminal angular momenta predicted by *MCRM* and *CCRM* is caused by the term $\bar{\omega}_+ \cdot T$ in the *MCRM* Eq. (39), which is not in the *CCRM* Eq. (4), resulting in different cranked frequencies, lower angular velocity, and hence higher terminal angular momentum predicted by the *MCRM* (as noted in Section 3). Including more realistic interaction, particularly pairing, in the model may reduce the terminal angular momentum predicted by the *MCRM*.

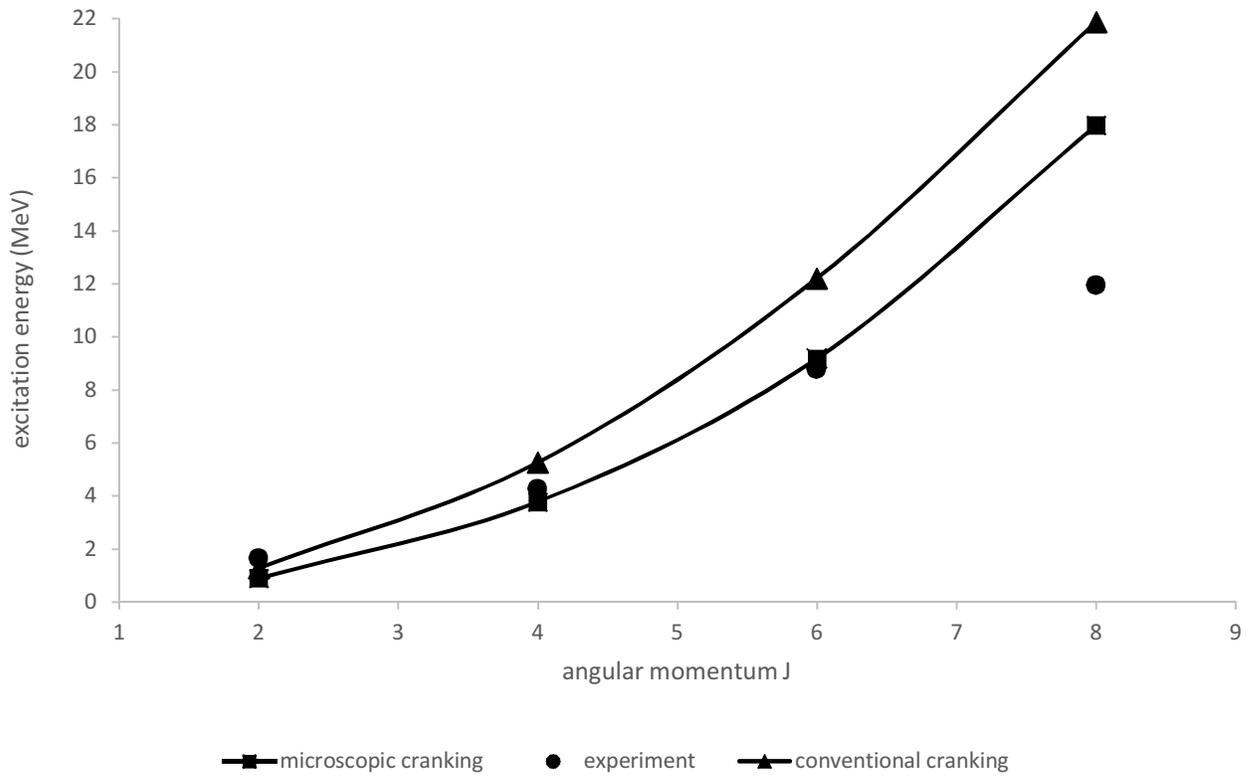

Fig 1: excitation energy versus J

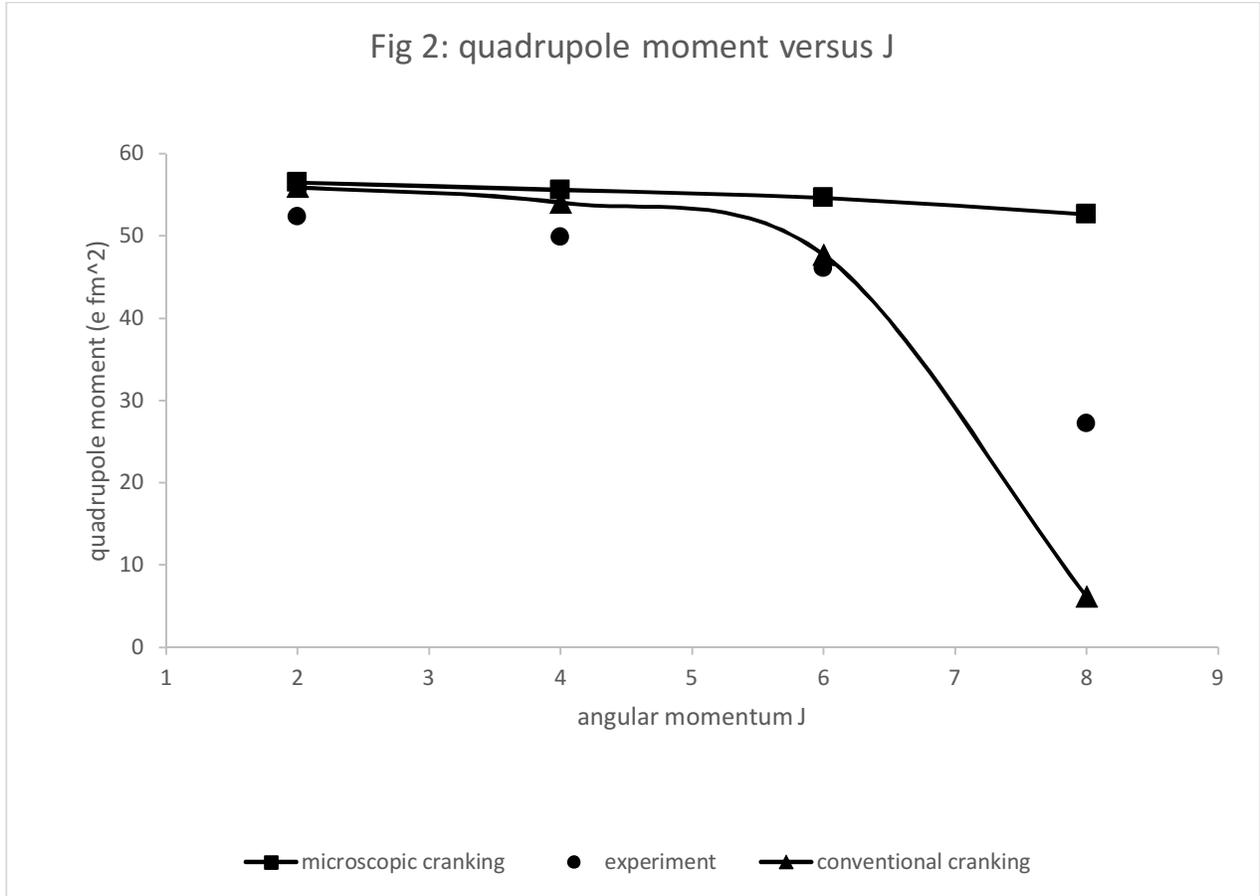

Fig 2: quadrupole moment versus J

## 5. Concluding remarks

Because of the success of the often-used conventional cranking model (*CCRM*) in the study of rotational properties of deformed nuclei, it behoves us to seek a microscopic understanding of the model. Among the many other attempts at this objective, the rigid-irrotational flow transformation in the previous microscopic cranking model (*MCRM*) for nuclear collective rotation about a single axis and its coupling to intrinsic motion is generalized in this article. This generalization allows us to consider the limit of vanishingly small collective angular velocity and angular momentum. In this limit, the angular velocity associated with Coriolis energy term in the *MCRM* Schrodinger equation for the intrinsic wavefunction becomes independent of the collective angular velocity, and remains finite, and is determined by the angular momentum constraint on the intrinsic wavefunction. In this limit, the time-invariance of the *MCRM* equation is destroyed. In the limit of vanishingly small collective angular velocity, the collective flows are large and oppose and cancel one another. In this limit, the resulting generalized-*MCRM* equation for the angular momentum constraint on the intrinsic wavefunction reduces to that of the *CCRM*, and the *MCRM* Schrodinger equation reduces to that of the *CCRM* except for an added irrotational-flow Coriolis kinetic energy term.

For a simple deformed harmonic oscillator potential, the two *MCRM* equations are solved analytically for the cranked frequencies and energy eigenvalues subject to the constant nuclear volume condition. The results are used to predict the ground-state excitation energy ($\Delta E_J$) and quadrupole moment ($Q_o$) in $^{20}_{10}Ne$. The *MCRM* predicts lower $\Delta E_J$ at and below $J = 4$ and higher $\Delta E_J$ above $J = 4$, particularly at $J = 8$. This behaviour is assumed to be caused by the neglect of pairing interaction in the model. At low $J$ values, pairing interaction couples pairs of nucleons to zero angular momentum reducing the moment of inertia and hence increasing $\Delta E_J$. At higher $J$ values, Coriolis force tends to align the quasi-particle angular momenta along the rotation eventually breaking up the paired nucleons resulting in higher moment of inertia and hence lower $\Delta E_J$. The *MCRM* predicts decreasing $Q_o$ with $J$ as in the experiment. However, the predicted $Q_o$ decreases sharply at the band cut-off angular momentum $J = 10$ instead of $J = 8$ in the experiment. It is assumed that including pairing interaction in the model may reduce the cut-off angular momentum.

Because of the presence of the irrotational-flow Coriolis energy term in the *MCRM* Schrodinger equation, the *MCRM* predicts cranked frequencies somewhat different than those predicted by the *CCRM*. For this reason, the *MCRM* predicts lower angular velocity and hence $\Delta E_J$, and higher band cut-off angular momentum and hence $Q_o$ than those predicted by the *CCRM*.

In a future article, we intend to study the impact of pairing interaction on the above *MCRM* results.